\documentclass[12pt,a4paper]{article}
\usepackage{amsmath}
\usepackage{amssymb}
\usepackage{graphicx}
\usepackage{epsfig}
\usepackage{psfrag}

\newcommand{\av}[1]{\langle #1 \rangle}
\newcommand{\xv}{{\bf x}}

\newcommand{\tr}{\operatorname{Tr}}
\newcommand{\Tr}{\operatorname{Tr}}
\newcommand{\re}{\operatorname{Re}}
\newcommand{\im}{\operatorname{Im}}

\newcommand{\plx}{v(\xv)}
\newcommand{\pl}{\mathcal{L}}
\newcommand{\nc}{\newcommand}

\nc{\be}{\begin{equation}}
\nc{\ee}{\end{equation}}
\nc{\bea}{\begin{eqnarray}}
\nc{\eea}{\end{eqnarray}}
\nc{\wlambda}{{{\widetilde\lambda}}}
\nc{\hlambda}{{{\widehat\lambda}}}
\nc{\hh}{{{\widehat{h}}}}

\nc{\cA}{{\cal A}}
\nc{\cB}{{\cal B}}
\nc{\cC}{{\cal C}}
\nc{\cD}{{\cal D}}
\nc{\cE}{{\cal E}}
\nc{\cF}{{\cal F}}
\nc{\cG}{{\cal G}}
\nc{\cH}{{\cal H}}
\nc{\cI}{{\cal I}}
\nc{\cJ}{{\cal J}}
\nc{\cK}{{\cal K}}
\nc{\cL}{{\cal L}}
\nc{\cM}{{\cal M}}
\nc{\cN}{{\cal N}}
\nc{\cO}{{\cal O}}
\nc{\cP}{{\cal P}}
\nc{\cQ}{{\cal Q}}
\nc{\cR}{{\cal R}}
\nc{\cS}{{\cal S}}
\nc{\cT}{{\cal T}}
\nc{\cU}{{\cal U}}
\nc{\cV}{{\cal V}}
\nc{\cW}{{\cal W}}
\nc{\cX}{{\cal X}}
\nc{\cY}{{\cal Y}}
\nc{\cZ}{{\cal Z}}
\nc{\bk}{{{\bf k}}}
\nc{\bx}{{{\bf x}}}

\nc{\simo}[1]{{\stackrel{#1}{\simeq}}}
\nc{\geqo}[1]{{\stackrel{#1}{\geq}}}
\nc{\geo}[1]{{\stackrel{#1}{>}}}
\nc{\guo}[1]{{\stackrel{#1}{\succ}}}

\nc{\Eq}{Eq.~}
\nc{\nr}[1]{(\ref{#1})}

\begin{document}

\begin{center}
\large{\bf A Gauge Theory of Wilson Lines as a Dimensionally
Reduced Model of $QCD_3$}
\end{center}

\begin{center}
P. Bialas$^{1,a}$, A. Morel$^{2,b}$, B. Petersson$^{3,c}$

\end{center}

\centerline{$^{1}$Inst. of Physics, Jagellonian University}
\centerline{ul. Reymonta 4, 33-059 Krakow, Poland}
\vspace{0.3cm}
\centerline{$^{2}$Service de Physique Th\'eorique de Saclay, CE-Saclay,}
\centerline{F-91191 Gif-sur-Yvette Cedex, France}
\vspace{0.3cm}
\centerline{$^{3}$Fakult\"at f\"ur Physik, Universit\"at Bielefeld}
\centerline{P.O.Box 100131, D-33501 Bielefeld, Germany}
\vspace{0.3cm}

\vspace{0.3cm}

\begin{abstract} \normalsize

We analyze a two dimensional
SU(3) gauge model of Wilson lines as a
dimensionally reduced model of high
temperature $QCD_3$. In contrast to perturbative dimensional
reduction it has an explicit global $Z_3$ symmetry in the
action. The phase diagram of the model is studied in the space of 
two free parameters used to describe the self interaction of the 
Wilson lines. In addition to the confinement-deconfinement transition,
the model also exhibits a new $Z_3$-breaking phase. These findings are
obtained by numerical simulations, and supported by a perturbative 
calculation to one loop.  A screening mass
from Polyakov loop  correlations is calculated numerically. It matches 
the known $QCD_3$ mass in a domain of parameters belonging to the normal 
deconfined phase.  
\end{abstract}

\vfill

\noindent
$^a$pbialas@th.if.uj.edu.pl \\
$^b$morel@spht.saclay.cea.fr \\
$^c$bengt@physik.uni-bielefeld.de \\

\newpage

\section{Introduction}

It would be of considerable interest to understand the  long distance 
properties in the quark gluon plasma quantitatively. Direct simulations
of QCD are still too costly to be continued to the continuum limit, at
least as far as these properties are concerned. Dimensional reduction
has
been shown to be a reliable approximation, if the temperature is
high enough. In this approach, the non static modes of the fields are
integrated out perturbatively, in practice usually to one or two loop
order. The resulting effective theory of the static mode 
is a bosonic field theory in one dimension less  
\cite{ginsparg}--\cite{thomas}. This becomes a very powerful technique, when 
the long distance properties are 
calculated non perturbatively on the lattice in the much simpler
reduced theory \cite{lacock}--\cite{us1}.
In general one finds that if the temperature is larger than about
twice $T_c$, the agreement with the full theory in pure gauge theories, 
where it can be checked, is good. A review of the situation can
be found in \cite{philipsen}.

In a series of articles \cite{us1,us2,us3}
 we have investigated the dimensional
reduction of SU(3) gauge theory in three dimensions to a two dimensional
Higgs-gauge-model. In this  case one can also easily simulate the full theory
and compare the results in detail. 
We found that using the one loop approximation
in the perturbative integration of the non static modes, the reduced
model
describes very well the correlations between Polyakov loops as well as
spacelike Wilson loops down to about $1.5$ times the critical
temperature. It should be emphazised that no free parameters are
involved, because the lowest order terms of the self-interaction
(potential) of the
Higgs field are determined by the perturbative calculation.

However, this reduced model does not have the $Z_3$ symmetry of the full model.
It therefore does not have the confinement-deconfinement phase transition.
As we pointed out in \cite{us4} 
one may construct a model,
which has the perturbatively reduced model as the high temperature limit, but
which does not break the $Z_3$ symmetry. Similar models have been
discussed in \cite{banks,pisarski}.

In this paper we investigate a model where the basic degrees
of freedom are the spatial components of the gauge field and the 
Wilson lines of the three dimensional pure gauge theory. The 
traces of the latter are the Polyakov loops, order parameters for the
preserved $Z_3$-symmetry. The Polyakov
loop potential is left undetermined and 
we parametrize it by the first two terms 
of a small loop expansion. We study the phase diagram of this model in the
corresponding parameter plane and show
that in addition to the expected confined and deconfined phases, 
another $Z_3$-breaking phase, distinct from the normal deconfined phase, 
appears.

The Polyakov loops correlation length is also calculated at high
temperature along the line 
corresponding to a purely quadratic potential. It matches that of 
the original 3D gauge theory at a point which lies inside 
the {\it normal} deconfined phase, not the new one. In this region of 
parameters,
it is thus consistent to view the model as a dimensionally reduced model 
for $QCD_3$, although an effective gain in computing cost 
requires an a priori knowledge of the Polyakov loops potential, not
calculated in this paper.  

These calculations are
performed via Monte Carlo simulations. First
accounts of the numerical results have been presented in
\cite{proc3,nara}. Furthermore we show analytically that
the one loop approximation gives a qualitative description
of the phase diagram. Although our numerical simulations are restricted
to the two dimensional reduced model, the one loop approximation can 
also be directly applied to the three dimensional case, which is the 
reduced model
of the physical four dimensional QCD. We see in this approximation the
same qualitative behaviour of the reduced model, as in the two
dimensional case, including the existence of the new phase.

In section 2 we define the reduced model, in section 3 we describe the
numerical simulations, and in section 4 we give the corresponding results.
In section 5 we derive the effective action 
of the reduced model in the one loop 
approximation, and compare with the results of the simulations. We also
present the analytic 
result in the corresponding three dimensional reduced model.
Finally section 6 contains
a summary of our results.

\section{The Model and its Relationship to Finite Temperature $QCD_3$}

In this section, we first write down the action of the 2D-model under
study and then discuss under which conditions it can be viewed as a
dimensionally reduced action for the Polyakov loops of pure QCD in 3
dimensions.

The model is defined by its partition function on a square
$L^2$ lattice with sites $\bx$, spanned by unit
vectors $\widehat i$, i=1,2. The spacing $a$ will be usually set to 1, unless
necessary when dimensionful quantities are defined. All dynamical variables 
belongs to the SU(3) group, gauge fields $U(\bx,i)$ sitting on the links, 
and matter fields $V(\bx)$, localized on the sites. The variables $V(\bx)$ 
will be later identified with the Wilson lines of $QCD_3$, whose normalized 
traces are the Polyakov loops. The partition function is written

\bea 
   Z & = & \int \cD U\,\cD V \; 
   \exp{( -  S_U - S_{U,V} - S_{V} )} ,
   \label {partition} \\
S_U &=&  \beta_2 \sum_{\bx}\,
    \biggl(1 - \frac{1}{3} \Re\, \tr [U(\bx;1) U(\bx+\widehat 1;2)
           U^{\dagger}(\bx+\widehat 2;1) U^{\dagger}(\bx;2)]
               \biggr),  \label{su} \\
S_{U,V}&=&t\, \sum_{\bx} \sum_{i=1,2}
  \biggl(1-\frac{1}{3} \Re\, \tr [U(\bx;i)
            V(\bx+\widehat {i}) U^{\dagger}(\bx;i) V^\dagger(\bx)]
                  \biggr). \label{suv} 
\eea

The measure $\cD U\,\cD V$ represents the products of the Haar
measures associated with the $U$ and $V$ variables, $S_U$ is the 2D
Wilson action and $S_{U,V}$ a gauge invariant kinetic term for the
matter fields, defining lattice couplings $\beta_2$ and $t$. The
self-interaction $S_{V}$ is assumed to be a real and locally gauge
invariant function of $V$, to be specified later.

Actions of the above type have already been considered in 
\cite{us4,banks,pisarski} as representative of what an effective
action for the Wilson lines of $QCD_3$ might look like. A first
account of some of our results has been presented in 
\cite{proc3, nara}. 

The connection of the partition function (\ref{partition}) to $QCD_3$ may be
thought of as follows. Adding a third dimension (a time direction 0), of length 
$a\,L_0$, to the previous lattice, $QCD_3$ at temperature $T=1/(a\,L_0)$ is 
defined by the 3-dimensional Wilson action on this new lattice and a 
coupling $\beta_3$:
\begin{eqnarray}
S_{QCD_3} &=& \beta_3 \sum_{x_0=1}^{L_0}\,\sum_{\bx}\, \sum_{\mu<\nu=0}^2
\nonumber \\
&&\biggl( 1 - \frac{1}{3} \Re \,\tr \,U(x;\mu)\, U(x+\widehat\mu;\nu)
\,U(x+\widehat\nu;\mu)^\dagger \,U(x;\nu)^\dagger
\biggr). \label{qcd3}
\end{eqnarray}

The continuum limit is the limit $a\to 0$ when
\begin{equation}
g_3^2=\frac{6}{a\,\beta_3}, \label{g3}
\end{equation}
which sets the energy scale, is kept fixed. So we define a dimensionless
temperature by 
\begin{equation}
\frac{T}{g_3^2}=\frac{\beta_3}{6\,L_0}, \label{redtemp}
\end{equation}
and fixing the temperature amounts to letting $\beta_3$ and $L_0$ go to
$\infty$ in a fixed ratio. Of course at the same time, the spatial lattice size 
$a\,L$ is supposed to be large compared to any correlation length.
Dimensional reduction \cite{ginsparg,appelquist} provides a way to 
extract properties of $QCD$ in the large $T/g_3^2$ regime. The application 
of the general procedure \cite{nadkarni,landsman,thomas} to 
$QCD_3$ is described in details in \cite{us1}.

Here we only recall the main steps of the process. 
In the Wilson action (\ref{qcd3}), the dynamical field variables
$U(x;\mu) \in SU(3)$ are related to the gauge fields by 
\begin{equation}
U(x;\mu)\equiv \exp[i\,A_{\mu}(x)]. \label{umu}
\end{equation}
To make the reduction, we choose a static gauge  
\begin{equation}
A_0(x_0,\bx)=A_0(\bx),
\end{equation}
which means
\begin{equation}
U(x_0,\bx;0)=U(\bx;0)
\end{equation}
for all $x_0$. The variables $V(\bx)$, which we call Wilson lines, are
defined in this gauge by
\begin{equation}
V(\bx)= U^{L_0}(\bx;0)=\exp[i\,L_0\,A_0(\bx)],  \label{polyakov}\\
\end{equation}
and are static operators. Their normalized traces are the gauge
invariant Polyakov loops, denoted $v(\bx)$:
\begin{equation}
v(\bx)=\frac{1}{3}\tr\,V(\bx).  \label{ploop}
\end{equation}
The spacelike fields $A_i(x_0,\bx)$, $i=1,2\,,$ can be split
into static and non-static components in this gauge,
\bea 
A_i^s(\bx)&\equiv&\frac{1}{L_0}\sum_{x_0}A_i(x_0,\bx), \label{as} \\
A_i^{ns}(x_0,\bx)&\equiv&A_i(x_0,\bx)-A_i^s(\bx). \label{ans}
\eea
The last step of standard dimensional reduction consists in 
integrating perturbatively over the $A_i^{ns}$ fields. For this purpose,
the lattice action (\ref{qcd3}) is expanded to an appropriate order in powers 
of $A_0$ and $A_i$, and full gauge fixing is achieved \cite{curci, STALG} 
by setting  the condition
\begin{equation}
\frac{1}{L_0}\,\sum_{x_0} \sum_{i=1,2} \,
\biggl (A_i(x_0,\bx)-A_i(x_0,\bx-\widehat i)\biggr )=0.
\end{equation}
Together with Eqs. (\ref{as},\ref{ans}), it implies 
\begin{equation} \label{landau}
\sum_{i=1,2}\,\biggl (A_i^s(\bx)-A_i^s(\bx-\widehat i)\biggr )=0,
\end{equation}
which is a lattice Landau gauge for the 2D gauge fields $A_i^s(\bx)$.
By integrating perturbatively over the gauge fields $A_i^{ns}(x_0,\bx)$
up to one loop order, one gets a two dimensional effective gauge theory
of the static modes, where $A_i^s(\bx)$, $i=1,2\,,$ are the gauge fields
and $A_0(\bx)$ acts like a Higgs field in the adjoint representation.
There are no infrared divergences in this perturbative calculation, and
one obtains at a given order in $g_3^2/T$ only a finite number of
interaction terms. The effective two dimensional model has, however,
severe infrared divergences in perturbation theory and thus has to be
treated non perturbatively. The lattice version of the reduced action
is

\bea
&&S_{2L}=S_{U}+S_{U,A_0}+S_{A_0}, \label{s}\\
&&S_U = \beta_3 L_0 \sum_\bx
\biggl( 1 - \frac{1}{3} \Re\, \tr U(\bx;1)
U(\bx+\widehat 1;2)
U^{\dagger}(\bx +\widehat 2;1) U^{\dagger}(\bx;2)
\biggr), \nonumber\\
&&S_{U,A_0} =\frac {\beta_3 L_0}{6}\sum_{\bx}
\sum_{i=1}^{2}
\tr \biggl( D_{i}(U) A_0 (\bx) \biggr)^2, \label{sua0}\\
&&D_{i}(U) A_0 (\bx) = U(\bx;i) A_0 (\bx+
\widehat i) U^{\dagger}(\bx;i) - A_0(\bx),
\nonumber \\
&&S_{A_0} = \sum_{\bx}
k_2\,\tr A_0 ^2(\bx)
+ k_4 \left( \tr A_0^2(\bx)
\right)^2. \label{sa0}
\eea
In the above formulae, $S_U$ is the pure gauge term in 2D,
$S_{U,A_0}$ the gauge invariant kinetic term for the scalar Higgs
field and $S_{A_0}$ the scalar potential, whose self couplings
$k_2$ and $k_4$ result from the one loop integration over the
non-static components of the gauge fields \cite{us1}. All terms
have a global $R_\tau$ symmetry $A_0(\bx)\,\to \,-A_0(\bx)$,
while the $Z_3$ symmetry of the original 2+1 dimensional SU(3)
is broken by the perturbative reduction procedure.

As shown in \cite{us1}, the 2D model defined by  the lattice
action (\ref{s}) successfully accounts for non-perturbative 
properties of $QCD_3$, at $T>1.5\,T_c$ and large distances, a  
conclusion reached from the behaviour of Polyakov loops correlators.

We now examine under which
conditions the model of Eq.(\ref{partition}) may also be related 
to this perturbatively reduced model. First the pure gauge piece $S_U$ of 
Eq.(\ref{su}) coincides with that of $S_{2L}$ if
\begin{equation}
\beta_2=L_0\,\beta_3. \label{beta2}
\end{equation}
Next consider the content of $S_{U,V}$, Eq.(\ref{suv}), for $V$ 
given by (\ref{polyakov}) and $A_0$ small, the situation in
which dimensional reduction is known to be justified. To second 
order in $A_0$, Eq.(\ref{suv}) generates the lattice invariant kinetic
term 
\begin{equation}
\frac{t\,L_0^2}{6}\, \sum_{\bx} \sum_{i=1,2}
  \Tr \biggl [A_0^2(\bx) +A_0^2(\bx+\widehat {i})
  -2\,U(\bx;i) A_0(\bx+\widehat {i}) U^{\dagger}(\bx;i) A_0(\bx)
          \biggr], \label{kina0} 
\end{equation}
which coincides with that given by Eq. (\ref{sua0})  provided one chooses
\begin{equation}
t=\beta_3/L_0.
\end{equation}
In other words, the coupling $t$ of the model is proportional to the
dimensionless temperature defined by Eq. (\ref{redtemp}).

We finally come to the last term $S_V$ in (\ref{partition}), 
unspecified yet. Of course one would like to be able to compute it from
$QCD_3$, as it is the case for the  self interaction $S_{A_0}$,
whose lowest order terms (\ref{sa0}) were computed
perturbatively in the high $T$ and large distances situation \cite{us1}.
In the absence of a known scheme to do so, we choose the simple form
\begin{equation}
S_V=\sum_\bx\,\biggl (\lambda_2\,|v(\bx)|^2 \label{sv}
    + \lambda_3\,\Re v^3(\bx) \biggr ), 
\end{equation}          
which fulfills the following requirements. It is real and locally gauge
invariant. It is also invariant under the global $Z_3$ transformation
\bea \label{z3}
V(\bx)\, &\to& z_n\,V(\bx), \\
z_n&=&\exp(2\,i\,\pi\,n/3)\,;\quad n=0,1,2\;mod(3). \nonumber
\eea
This in fact true for the full action of our model, and distinguishes it
from perturbative dimensional reduction.
Being an SU(3) matrix, $V(\bx)$ possesses two (invariant) degrees of freedom 
only, for example two of its three eigenvalues 
\bea
V_\alpha (\bx) \equiv \exp\biggl[i\,\theta_\alpha (\bx)\biggr ],
\label{eigenv}  \\
\prod _{\alpha =1}^{3}\,V_\alpha (\bx) =1,
\eea
or equivalently the real and imaginary parts of its normalized trace $v(\bx)$, 
an order parameter for the $Z_3$ symmetry. A self-interaction like
(\ref{sv}) opens the possibility of a transition from a symmetric phase,
where $\av {v(\bx)}$ vanishes, to a broken phase where it has a finite
modulus. From this point of view, we consider (\ref{sv}) as the first 
terms of an expansion  of $S_V$ near the phase transition. The
attempt made in \cite{us4} was to fix $\lambda_2$ and
$\lambda_3$ in such a way that, for $A_0$ small, $S_V$ matches the two 
lowest order terms of the expansion (\ref{sa0}) of $S_{A_0}$.

The use of free parameters in $S_V$ makes the model less predictive than
perturbative reduction at high temperature. In turn, a
parametrization like (\ref{sv}) allows to investigate which
non-perturbative effects might be associated with lower temperatures.
This is why, in what follows, we focuss on the structure of the phase diagram,
which we study numerically in the
$\{\lambda_2,\beta_3\}$ plane at $\lambda_3=0$, the effect of a
third order term in $v$ being discussed at a perturbative
level only  (section 5).

\section{Numerical simulations}

The model \eqref{partition} was studied using conventional lattice QCD
 Monte-Carlo techniques. Lattices of sizes ranging from $L^2=16^2 $ to 
 $72^2$ were studied and the parameter $L_0$ (the inverse
 temperature of the $2+1$ model in lattice units) was always set to $4$.
 The multi hit metropolis algorithm
(with 8 hits) was used for updating both the gauge fields $U(\bx,i)$ and
the Wilson lines $V(\bx)$.  
The dynamics of the two sectors were quite different: While the
integrated autocorrelation time for the plaquette operator was
negligible (of the order of 10 sweeps), for the Wilson lines it could
rise up to 25000 sweeps near the phase transition on a $32^2$
lattice. One sweep consisted of one update of all the $U$ fields followed by 
one update of all the $V$ fields. Typically, runs with 2-4 $10^6$ sweeps 
were performed for the system on a $32^2$ lattice around the phase transitions.

The quantities measured concerned the Polyakov loops $v(\bx)$ defined
in Eq. (\ref{ploop}): their lattice averages
\begin{equation}
\pl=\frac{1}{L^2}\sum_\xv \plx,
\end{equation}
their two--point on axis correlators
\begin{equation}\label{eq:LL}
P(r)=\frac{1}{L^2}\av{\sum_{x_1}\sum_{x_2}v(x_1,x_2)v^*(x_1+r,x_2)}
\end{equation}
and projected correlators~: 
\begin{equation}\label{eq:LLprj}
P_{prj}(r)=
\frac{1}{L^2}
\av{\sum_{x_1}\sum_{x_2} v(x_1,x_2)\sum_{x'_2} v^*(x_1+r,x'_2)}.
\end{equation}

The traces $\pl$  were measured and
stored on a per configuration basis, so that their distribution $P(\pl)$
and the susceptibility
\begin{equation}\label{eq:xi}
\chi=\av{|\pl|^2}-\av{|\pl|}^2
\end{equation}
could be obtained.  

The correlators were fitted with the appropriate formulas (see 
section 4.2) using correlated fit procedures in order to extract
screening masses \cite{cfit}. 

All the errors were calculated using a blocked bootstrap method.  The
data were divided into blocks,  the size of which was adjusted so as
to make them independent from each other, but at least
ten blocks were used. Then a bootstrap sample was drawn from those
blocks and the measurements performed on this sample.  The
procedure was repeated 200 times and the standard deviation of the
resulting distribution of measured values taken as 
an estimate of the error on their average.

\section{Results from Numerical Simulations}
The numerical simulations gave us evidence for the phase diagram 
in Fig.~\ref{fig:pd}, as we explain below in 4.1. Further in 4.2, we 
discuss the behaviour of the correlation length measured in the various phases.

\subsection{Phase structure}
 \begin{figure}
\begin{center}
\includegraphics[width=10cm]{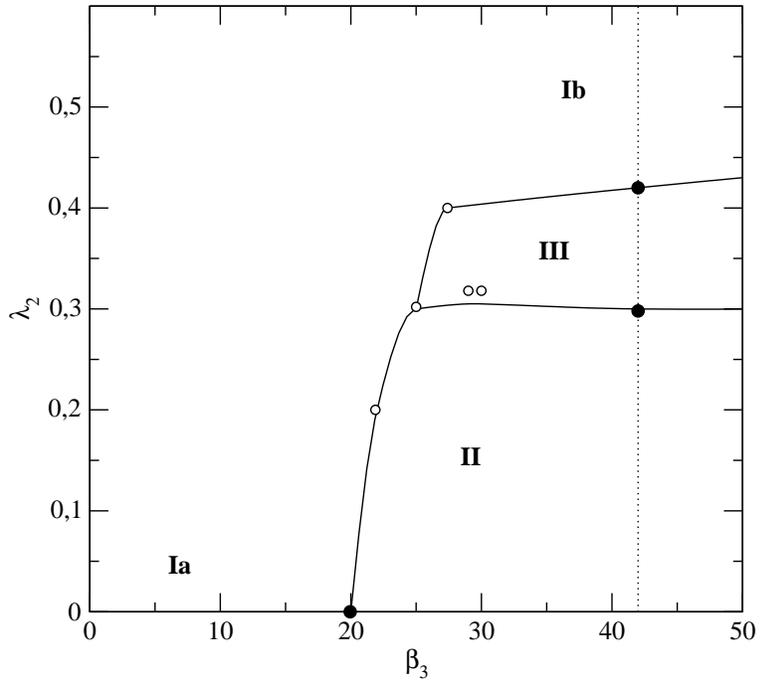}
\end{center}
\caption{\label{fig:pd}
  A tentative phase diagram. Open points denote the location of peaks
  in susceptibility observed on $16^2$ lattices. Black points
  denote the position of phase transitions established on $32^2$
  lattices with large statistics.  The lines are hand drawn sketches
  of where transitions take place.}
\end{figure}

\begin{figure}
\begin{center}
\psfrag{L}[][][.75][0]{$\av{|\mathcal{L}|}$}
\includegraphics[width=11cm]{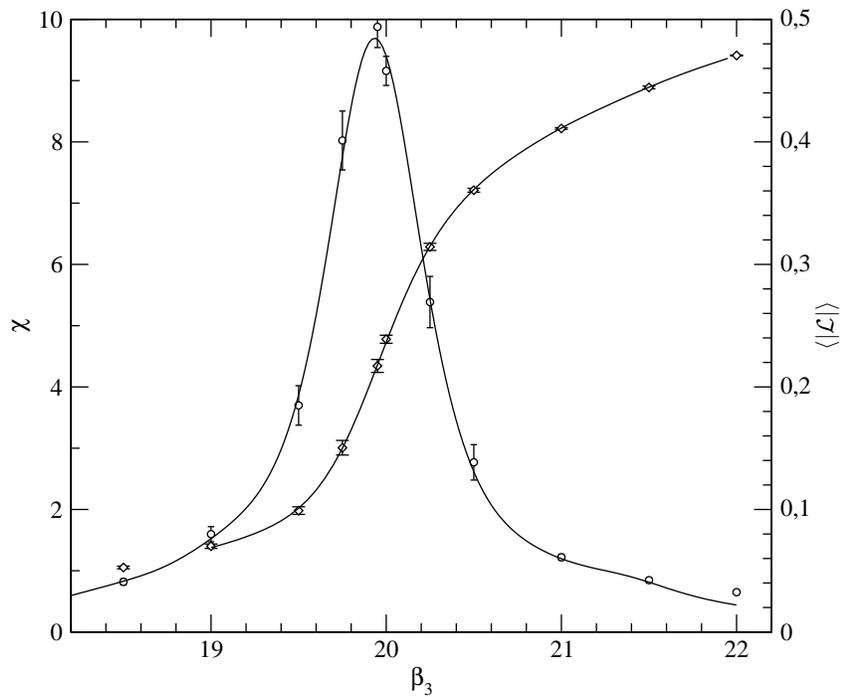}
\end{center}
\caption{\label{fig:xi_l20}Polyakov loop susceptibility (circles) and average
modulus (diamonds) versus $\beta_3$ for $\lambda_2=\lambda_3 =0$ on $32^2$ 
lattices. The continuous lines result from Ferrenberg--Swendsen reweighting.}
\end{figure}

We started with simulations of the so-called `naive' model, i.e. with no
self-interaction of the Polyakov loops: $S_V=0$ in
Eq.(\ref{partition}), that is $\lambda_2=\lambda_3=0$ with the parametrization
(\ref{sv}). There the model 
may also be viewed as standard 2+1 QCD on a lattice with one time slice only.
Exploring a range of $\beta_3$ values we observed a clear 
signal for the phase transition expected between a low temperature (low
$\beta_3$) confined $Z(3)$ symmetric phase (region Ia in Fig.~\ref{fig:pd}) 
and a broken symmetry high temperature phase (region II).The
corresponding peak in the susceptibility occurs at $\beta_3$ very close 
to 20 as shown in Fig.~\ref{fig:xi_l20}, where the average of $|\cL|$ is
also shown.

Still keeping $\lambda_3=0$, we then performed series of scans in the 
$\beta_3$--$\lambda_2$ plane on $16^2$ lattices, looking for peaks in the
susceptibility (\ref{eq:xi}) as signals for phase transitions and 
characterizing different phases by the corresponding distributions 
of  $\pl$. Although it also includes informations from larger lattices,
the tentative phase diagram of Fig.~\ref{fig:pd} results from this
exploration.

\begin{figure}
\begin{center}
\includegraphics[width=11cm]{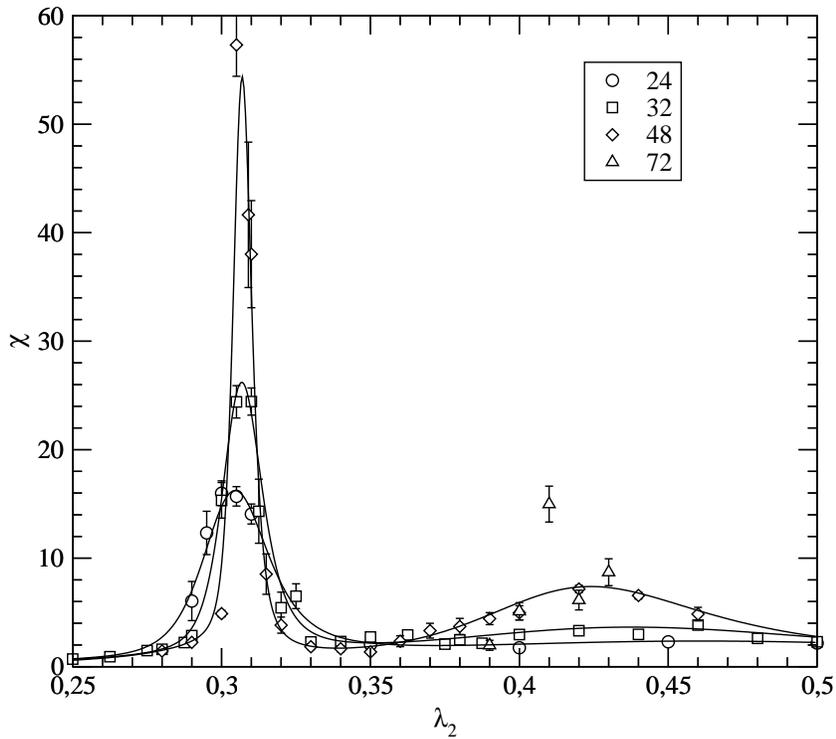}
\end{center}
\caption{\label{fig:xi}Polyakov loop susceptibility versus 
$\lambda_2$ for $\beta_3=42$ and $\lambda_3 =0$  on  lattices of various sizes.
The continuous lines result from Ferrenberg--Swendsen reweightings for the 
three smallest lattices.}
\end{figure}

\begin{figure}
\begin{center}
\includegraphics[width=13cm]{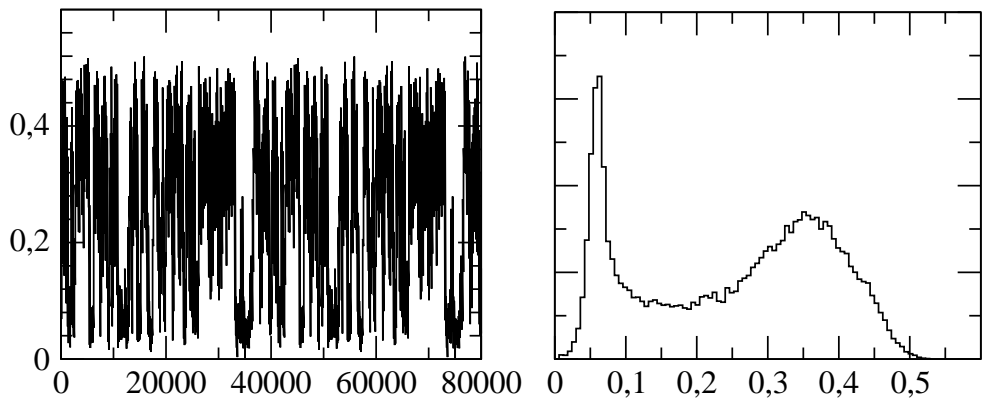}
\end{center}
\caption{\label{fig:fo}Time history and histogram for the quantity 
$|\pl|^2$, on a $32^2$ lattice for $\beta_3=42$, $\lambda_2= 0.305$
and $\lambda_3=0$. One clearly sees jumps between two distinct values, and the 
corresponding double peak structure of the histogram.}
\end{figure}

More precise scans in $\lambda_2$ along 
the line $\beta_3=42$ ($T\sim 2\,T_c$) give for the susceptibility the results
plotted in Fig.~\ref{fig:xi}.  
A strong transition signal shows up just above $\lambda_2=0.3$. There, the time 
history of $|\pl|^2$ and the corresponding histogram (Fig.~\ref{fig:fo}) 
clearly favour a first order phase transition. The height of the peak 
is furthermore compatible with a $L^2$ size dependence in the range
$L=24$--$48$, which is the finite size scaling expected for a
first order transition.

The evidence, from similar considerations, for a second phase transition 
around $\lambda_2 =0.4$ is considerably weaker.
A peak is hardly visible on the same lattices and we had to go to
$72^2$ lattices (with correspondingly poorer statistics) to ensure that
the bump detected does grow with the size. As we will immediately see however, 
the combination of the above observation with properties of the Polyakov
loop distributions confirms the existence of a second transition.
Furthermore, the theoretical analysis developed in the next section supports
the same conclusion.

\begin{figure}
\begin{center}
\psfrag{reL}[][][.7][0]{$\re\mathcal{L}$}
\psfrag{imL}[][][.7][0]{$\im\mathcal{L}$}
\includegraphics[width=14.5cm]{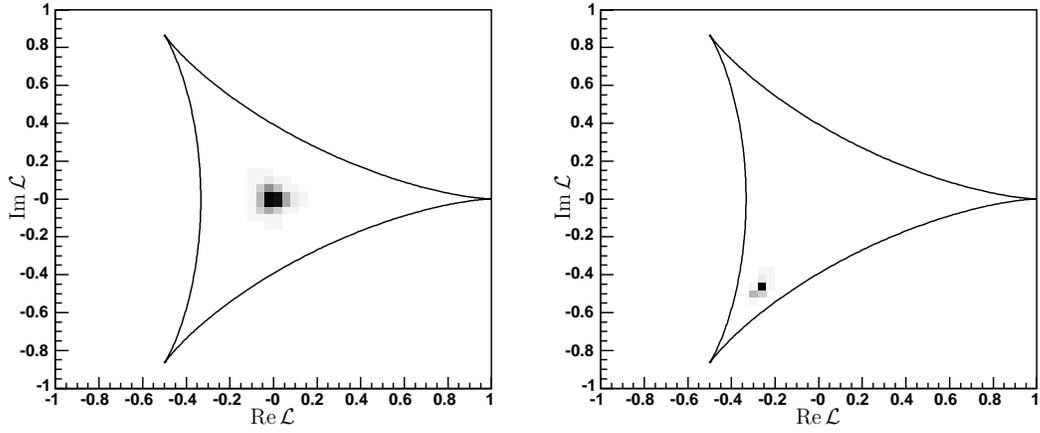}
\end{center}
\caption{\label{fig:hist-I-II}Typical distributions of $\pl$ at
$\lambda_2=\lambda_3=0$ for phases Ia
($\beta_3=18.0$, left) and II ($\beta_3=24$, right).}
\end{figure}

In order to identify the nature of the various phases expected from the diagram 
of Fig.~\ref{fig:pd}, we looked at the corresponding distributions of
Polyakov loops. The phases labelled Ia and II  
are easily identified with standard symmetric and broken 
$Z(3)$ phases respectively. In
Fig.~\ref{fig:hist-I-II}, these distributions are shown in
the complex plane of $\pl$, and they exhibit the expected patterns:
The values of $\pl$ cluster around 0 in phase Ia, and have a finite
modulus and an argument of the form $2\,i\,\pi\,n/3$ in phase II.

\begin{figure}
\begin{center}
\psfrag{reL}[][][.7][0]{$\re\mathcal{L}$}
\psfrag{imL}[][][.7][0]{$\im\mathcal{L}$}
\includegraphics[width=14.5cm]{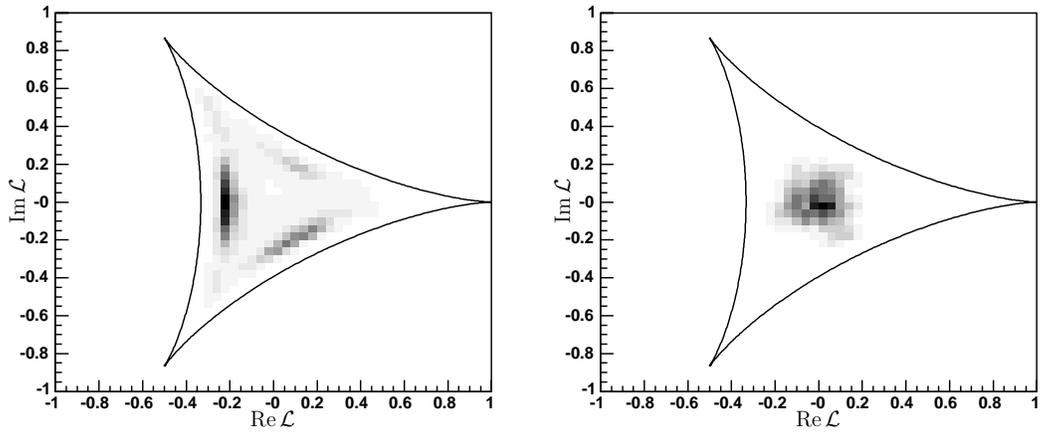}
\end{center}
\caption{\label{fig:hist-III-I}Typical distributions of $\pl$ at
$\beta_3 =42$ and $\lambda_3 =0$ for phases III
($\lambda_2 =0.32$, left) and Ib ($\lambda_2 =0.48$, right).}
\end{figure}

As we increase $\lambda_2$ from 0 at $\beta_3=42$ and move from region
II to region III, 
another symmetry breaking pattern appears. $Z_3$ symmetry breaking is
again manifested by clusters around non-zero values of $\pl$, but now 
arguments of the form $2\,i\,\pi\,(n+1/2)/3$ are favoured. 
This is illustrated on Fig.~\ref{fig:hist-III-I} (left). In this example,
the lattice is not large enough to prevent the system from tunneling
between different phases. Also the run is too short to allow for full
equilibration. But accumulations of values of $\pl$ close to the boundary with 
arguments around $\pi$ and $-\pi/3$ are clearly observed.
Increasing further $\lambda_2$, we move into a region Ib 
where $Z_3$-symmetry is recovered, as shown in Fig.~\ref{fig:hist-III-I}
(right). We conjecture that this phase
is the same as the low temperature confined phase (Ia).

\subsection{Screening masses}
\begin{figure}
\begin{center}
\includegraphics[width=11cm]{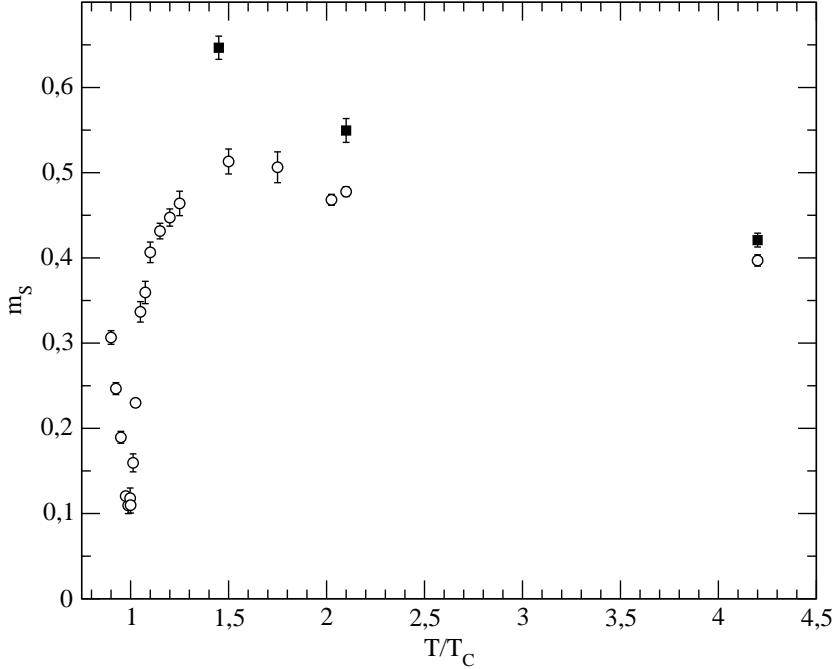}
\end{center}
\caption{\label{fig:massnp}
Screening masses in lattice units for the model (\ref{partition}) 
with $S_V=0$ (circles). At 
large $T$, they approach those obtained in naive dimensional reduction of 
$QCD_3$ (black squares, 2D adjoint Higgs model with no potential \cite{us1}).} 
\end{figure}

The inverse of the correlation length between Polyakov loops in the
present model corresponds to a screening mass in 2+1 QCD. This screening mass 
was obtained from  the two point Polyakov loop
correlators \eqref{eq:LL}  by fitting $m_S$ in
\begin{equation}
P(r)\approx 
a\biggl(\frac{e^{-m_S r}}{\sqrt{r}}+\frac{e^{-m_S (L-r)}}{\sqrt{L-r}}\biggr)+c.
\end{equation}
The constant $c$ accounts for the fact that we measured 
unconnected correlators. Fig.~\ref{fig:massnp} provides a comparison of $m_S$,
as obtained in the present model for $\lambda_2=\lambda_3=0$ 
(circles), with previous results (black squares) obtained in the naive 
perturbative 
reduction scheme (\cite{us1}, no Higgs potential). The data are
plotted versus the quantity $T/T_c \equiv \beta_3/\beta_{3c}$, 
where $\beta_{3c}$ is the critical value of $\beta_3$ for $L_0 =4$, 
respectively equal to 20 ( this model, see Fig.~\ref{fig:pd}) and to 14.7 in
2+1 QCD \cite{lego}. The two sets of points tend to join each other at
high temperature. It is the expected result since if $V$ becomes
equivalent to an element of $Z_3$, it can be rotated to the form
(\ref{polyakov}) with $A_0$ small, and the action in (\ref{partition}) can
thus be expanded in powers of $A_0$. In this situation, if there is no  
loop potential $S_V$ to generate a local Higgs potential, one ends
up with the naively reduced model. Near the transition, the two models
have very different behaviours. In the perturbative reduction there is
no $Z_3$ symmetry in the action, correspondingly no phase transition,
and the mass does not decrease near the $\beta_{3\,c}$ of the full
theory. In the present model, there is a phase transition and the mass
decrease from above the transition is consistent with a second order
transition. This in accord with the
behaviour of $QCD_3$.

\begin{figure}
\begin{center}
\includegraphics[width=11cm]{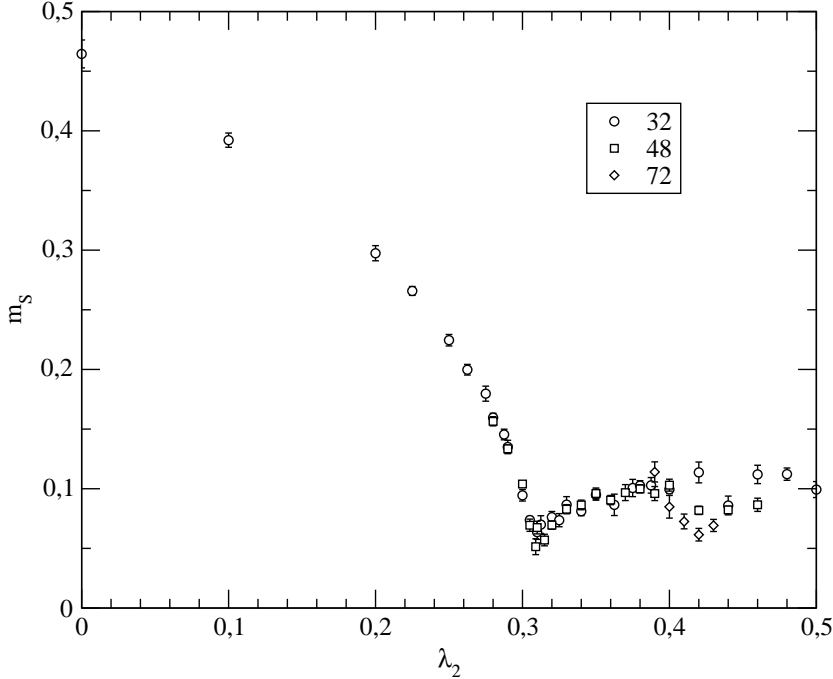}
\end{center}
\caption{\label{fig:mass}
The screening masses $m_S$ in lattice units, fitted with Eq.~\eqref{eq:LLprj} 
at $\beta_3=42$ and $\lambda_3=0$, as a function of $\lambda_2$ for various lattice sizes.}
\end{figure}

We also measured the screening masses as a function of $\lambda_2$ from
our data at $\beta_3 =42$ and $\lambda_3=0$. In this case, we used the 
projected (0--momentum) 
correlators \eqref{eq:LLprj} and extract the masses by fitting $m_S$ in
\begin{equation}
P_{prj}(r)\approx 
a\bigl(e^{-m_S r}+e^{-m_S (L-r)}\bigr)+c.
\end{equation}
Our results are summarized in Fig.~\ref{fig:mass}, for three lattice
sizes. Important size effects are observed in the region of the ($III\to
Ib$) transition just above $\lambda_2 =0.4$. It is not so near the
transition at $\lambda_2 =0.3$, which is consistent with evidences mentioned
earlier which favour a first order transition. It is important to
notice that the screening mass at $\beta_3$=42 (corresponding to
$T/T_c\simeq 2$) coincides with the screening mass of $QCD_3$ at
$\lambda_2=0.18$, that is {\it well inside} phase II. The situation is 
different from that in the perturbatively reduced model, where we found
the reduction point to be in a {\it metastable} $Z_2$ symmetric region
of parameter space, beyond the transition to the Higgs phase 
\cite{us1,philipsen}.

\section{Perturbative Approach to the Polyakov Loops Potential}

In order to get further insight into the phase structure of the model,
we have performed a perturbative calculation of 
the effective potential of the Polyakov loops. We assume that in a phase
where $Z_3$ is spontaneously broken, i.e. where $\av{v(\bx)}$ is not zero,
the Wilson line $V(\bx)$ is, up to a local gauge
transformation $\Gamma (\bx)$, a constant matrix which we denote $V$.
So any local fluctuation of $V(\bx)$ which cannot be gauged away is ignored,
and the Polyakov loop $v(\bx)$ is a constant number $v$ on the lattice:
\begin{eqnarray}
V(\bx)&=&\Gamma(\bx)\,V\,\Gamma ^\dagger(\bx),  \\
v(\bx)&=&\frac{1}{3} \tr V(\bx)=\frac{1}{3} \tr V=v.
\end{eqnarray}
Due to gauge invariance of both the action and the measure, $\Gamma (\bx)$ 
can then be absorbed in a redefinition of the group variables $U$.  
According to Eqs. (\ref{s}, \ref{sv}), the Polyakov loops potential 
$W_{P}(v)$ defined by
\begin{equation}\label{eq:WV}
\exp\bigl [-L^2 W_{P}(v)\bigr ]=\int\text{D}[U]\exp\bigl(-S\bigr), \nonumber
\end{equation}
is 
\begin{equation}
W_{P}(v)=\lambda_2 |v|^2 +\lambda_3 \Re v^3+W(v),
\end{equation}
where $W(v)$ is the contribution resulting from the coupling 
of $V$ to the gauge fields. In order to compute it to
one loop order, 
we first expand $S_U+S_{U,V}$ to second order in the SU(3)
algebra using
\begin{eqnarray} \label{afield}
U(\bx;i)&=&\exp[i\,g\,A_i(\bx)], \\
A_i(\bx)&=&\sum_{a=1}^{8} \,T^a\,A^a_i(\bx),\nonumber\\
\tr [T^a\,T^b]&=&\frac{1}{2}\delta_{ab}. \nonumber
\end{eqnarray}
Note that with respect to  previous definitions, $A_i(\bx)$ has been
rescaled by the gauge coupling $g$, where $g^2=6/\beta_2$. Including the
quadratic part of the jacobian from the $U\to\,A$ change of variables, we get 
\begin{eqnarray}
S^{(2)}_A=\frac{1}{2}\sum_{\bx}\Biggl [
           \,\sum_{a=1}^{8}  
   \Biggl(A^a_1(\bx)-A^a_1(\bx+\widehat{2})
         -A^a_2(\bx)+A^a_2(\bx+\widehat{1})\Biggr)^2 \nonumber \\
+\sum_{i=1}^2\Biggl (\frac{g^2\,t}{6}\sum_{a,b=1}^{8}A_i^a(\bx)A_i^b(\bx) 
   \biggl(2\delta_{ab}-M_{ab}-M_{ba} \biggr)+\frac{g^2}{4} 
\sum_{a=1}^{8}\,(A_i^a(\bx))^2\Biggr )\Biggr].\label{s2a} 
\end{eqnarray}
The $8\times 8$ constant matrix $M$ is defined by 
\begin{equation}
M_{ab}=2\,\tr\,[T^a\,V\,T^b\,V^\dagger], \label{mab}
\end{equation}
and its properties are  derived in the appendix.
Since integration over the $A_i$'s requires gauge fixing, we then 
add up to $S^{(2)}_A$ the $\xi$-gauge fixing term
\begin{equation}
S_{\xi}=\frac{1}{\xi}\,\sum_x\,\sum_a\,\biggl (A_1^a(\bx)-A_1^a(\bx-\widehat{1})
                +A_2^a(\bx)-A_2^a(\bx-\widehat{2}) \biggr )^2.
\end{equation}
Note that the limit $\xi\to \,0$ reproduces the Landau gauge
(\ref{landau}) in which the perturbative reduction was performed.
There is no contribution from the ghost terms to the one loop
potential. 
For the rest of the calculation, whose details are given in the
appendix, we go to Fourier space, using lattice momentum variables
\begin{equation}
k_i=\frac{2\pi n_i}{L},\quad \widehat k_i= 2 \sin k_i/2,
\quad \widehat k^2=\widehat k_1^2+\widehat k_2^2. 
\end{equation}
The quadratic form in $A_i$, diagonal in $k$-space, is a matrix $Q(k)$ in
the coordinate$\,\bigotimes\,$colour space so that 
\begin{equation}\label{eq:WU}
W(v)=
\frac{1}{2\,L^2}\, \sum_{k_1k_2}  \log \det Q(k)=
\frac{1}{2\,L^2}\, \sum_{k_1k_2} \Tr \log Q(k).
\end{equation}
Here, $\Tr$ denotes the trace over both the coordinate 
and color indices. We obtain 
\begin{equation}
Q_{ij}^{ab}=K_{ij}\delta_{ab}- 
\gamma^2\,M_{S,\,ab}\,\delta_{ij}, \label{qkm}
\end{equation}
where
\begin{eqnarray}
\gamma^2&=&\frac{1}{3}\,g^2\,t, \label{gamma2} \\
K_{ij}&=&(\widehat k^2+\gamma^2+\frac{g^2}{4})\,\delta_{ij}-
(1-\frac{1}{\xi})\,\widehat k_i\,\widehat k_j\,
\exp(i\frac{k_i-k_j}{2}),  \label{kij} \\
M_{S}&=& \frac{1}{2}\,(M+M^{T}). \label{ms}
\end{eqnarray}
The diagonalization of $K$ is trivial. That of $M_S$ is made in
the appendix, where in terms of the eigenvalues (\ref{eigenv}) of $V$, its 
eigenvalues $m_a$ are shown to be 
\begin{equation}
1,\,\, \cos(\theta_1-\theta_2),\,\,\cos(\theta_2-\theta_3),\,\,
\cos(\theta_3-\theta_1) \label{eigenm},
\end{equation}
each of them being doubly degenerate. Up to an additive
constant in $v$, the one loop contribution to the Polyakov loops
effective potential then reads 
\begin{eqnarray}
W(v)=\frac{1}{L^2}\sum_{k_1k_2}\,\,\sum_{\alpha<\beta =1}^3\, \log\, 
\biggl [\frac{1}{\xi}\,\biggl(\widehat k^2+\frac{g^2}{4}+2\,\gamma^2\,
\sin^2\,\frac{\theta _\alpha-\theta _\beta}{2}\biggr )
\,\times \nonumber\\
\biggl (\widehat k^2+\xi\,(\frac{g^2}{4}+2\,\gamma^2\,
\sin^2\,\frac{\theta _\alpha-\theta _\beta}{2})
\biggr )\biggr ]. \label{wv}
\end{eqnarray}
Because the eigenvalues $\exp(i\,\theta _\alpha)$ of $V$ are not independent,
it is more convenient to express $W$ explicitly as a function of the
Polyakov loop $v$, which is furthermore an order parameter for the 
$Z_3$ symmetry. This again is performed in the appendix.
Specializing to the Landau gauge
by taking the limit $\xi\rightarrow 0$ and dropping terms not depending on
$v$, we obtain 
\begin{eqnarray} 
W_P(v)&=&\lambda_2 |v|^2 +\lambda_3 \Re v^3+\frac{1}{L^2}\sum_{k_1k_2}\log\,
\Biggl [\biggl (\widehat k^2+\frac{g^2}{4}+\frac{3\gamma^2}{2}\biggr
)^3 \nonumber \\
&-&\frac{9\gamma^2}{2}
\biggl (\widehat k^2+\frac{g^2}{4}+\frac{3\gamma^2}{2}\biggr )
\biggl (\widehat k^2+\frac{g^2}{4}+3\,\gamma^2\biggr) |v|^2 \nonumber \\
&+&\frac{27\gamma^4}{2}\biggl (\widehat k^2+\frac{g^2}{4}+2\gamma^2\biggr
)\,\Re v^3
-\frac{81\gamma^6}{2^3}\, |v|^4 \Biggr ] .   \label{eq:wltrv}
\end{eqnarray}
Using this expression, we studied the 
shape in $v$ of $W_P(v)$ for $\beta_3=42$, $L_0=4$, $\lambda_3=0$ looking for 
its minimum as a function of $\lambda_2$.  

For $\lambda_2=0$ this minimum 
is reached at $|v|=1$, i.e. $v=\exp(2i\,n\,\pi/3)$, the standard
 $Z_3$-breaking phase II. Of course the situation
remains unchanged for any negative $\lambda_2$, which enhances the 
 Boltzmann weight.  On the contrary, large $|v|$ values are suppressed for 
 $\lambda_2$  positive, and the minimum is at $v=0$ for
 $\lambda_2$ large enough, this is phase Ib. There is however an interval
in $\lambda_2$ between $\simeq 0.25$ and $0.36$ where the minimum occurs
at the non trivial points $v=1/3\,\exp[2\,i\,(n+1/2)\,\pi/3]$, corresponding to
the phase III discovered in our numerical simulation 
inside a similar interval ($\lambda_2$ between $\simeq 0.3$ and
0.4, see Fig.~\ref{fig:xi}). The existence 
of this intermediate phase 
is due to the cubic term inside the $\log$ of $W(v)$ in Eq. (\ref{eq:wltrv}),
which is minimum for arg$(v^3)=\pi$.

\begin{figure} 
\begin{center}
\includegraphics[width=13.5cm]{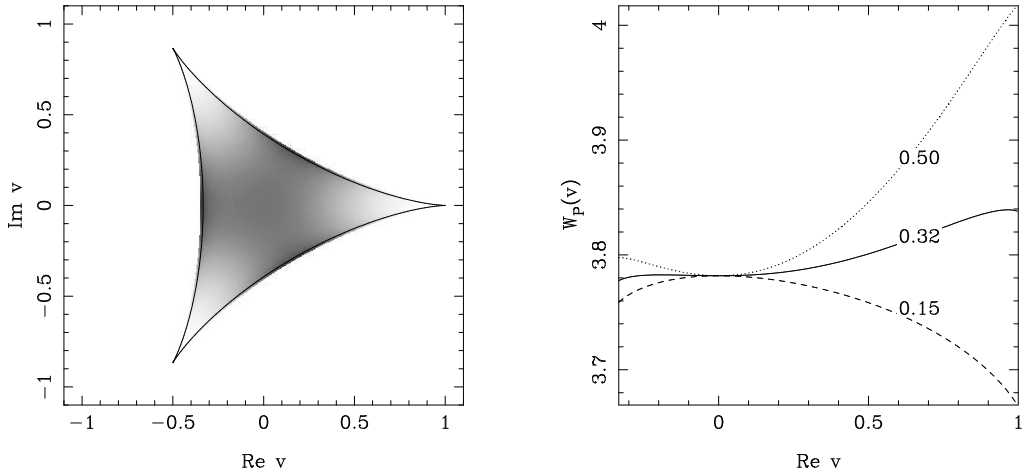} 
\end{center}
\caption{\label{fig:hv}
$\beta_3=42,\lambda_3=0$. On the left, the Polyakov loop potential $W_P(v)$ 
in the $v$-plane at $\lambda_2=0.32$. On the right, the shape of
$W_P(v)$ for $v$ real at three
values of $\lambda_2$. The positions of the minimum characterize phases II, 
III, I successively from bottom to top.}
\end{figure}

The situation just described is
illustrated in Fig. \ref{fig:hv}. On the left, we show $W_P(v)$ in the
complex $v$-plane for $\lambda_2=0.32$ inside the phase III, 
to be compared  with the similar
plot of Fig.~\ref{fig:hist-III-I} (left), the Boltzmann weight being the
largest (resp. smallest) in the blackiest (resp. clearest) regions. On
the right, $W_P(v)$ is plotted versus $v$ real, for $\lambda_3=0$ and three 
values of $\lambda_2$ corresponding to phases II, III, and Ib,
as seen from the locations of the minimum, respectively at $v$=1, $v$=--1/3 
and $v$=0.

\begin{figure} 
\begin{center}
  \includegraphics[width=11cm]{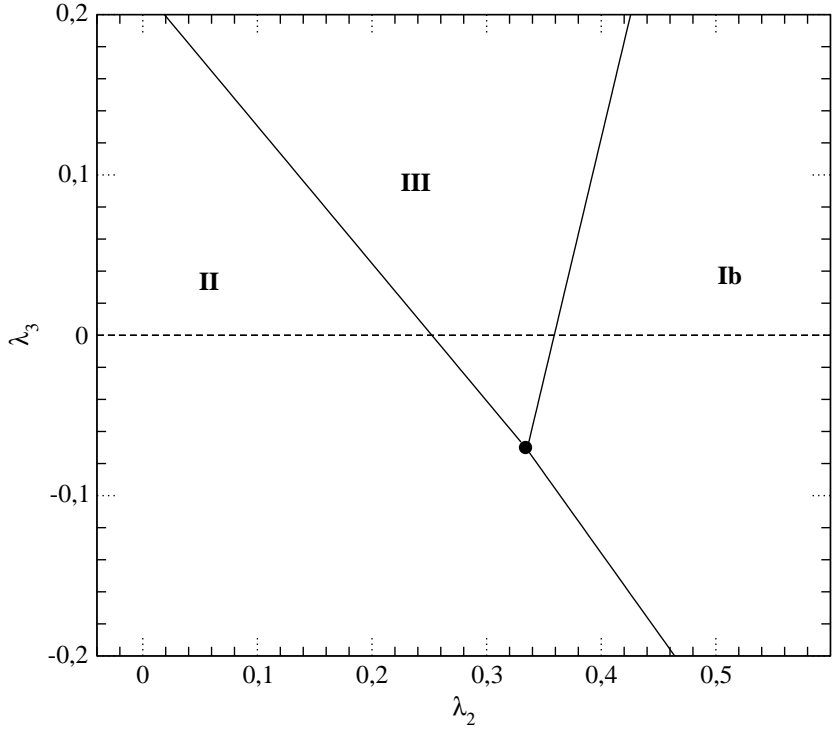}
\end{center}
\caption{\label{fig:l2l3pd}The phase diagram in the $\lambda_2 -
\lambda_3$ plane, as predicted from the one loop effective potential 
$W_P(V)$ at $\beta_3=42$. }
\end{figure}

The analysis can be extended to include the effect of
the term $\lambda_3 \Re v^3$. From the discussion above, it is clear that
a positive (resp. negative) value of $\lambda_3$ favours (resp. disfavours) 
phase III, which is characterized by $\Re v^3$ negative. In fact, 
computing $W_P(v)$ for $v$=--1/3, 0, 1 in the $\lambda_2$--$\lambda_3$ plane,
the perturbative phase diagram is easily obtained: At each point the true phase
corresponds to the lowest of the three $W_P$ values, a (straight)  transition
line occurs whenever the two lowest coincide, and a triple point when
all three are equal. The result is shown in Fig.~\ref{fig:l2l3pd}.

The lattice perturbative calculation leading to Eq. (\ref{wv}) is formally
similar to that made for SU(2) in Ref. \cite{weiss}. There, 
gaussian integration was performed over {\it both} the static and non static 
components of the $QCD_4$ spatial gauge fields. As a result of this purely
perturbative approach, the $Z_3$ symmetry could not be spontaneously
broken.

We end up this section with a few remarks. \\

{\bf Remark 1} 

The perturbative calculation of $W(V)$ can be easily extended to 
any dimension D.  The only change concerns the kinetic part $K_{ij}$
(\ref{kij}) whose indices now run from 1 to D. In the $\xi$-gauge, this 
K-matrix has $D-1$ eigenvalues equal to $K_1$ and one to $K_2$, 
where $K_1$ and $ K_2$ are given by Eq.~(\ref{eigenk}) of the appendix for 
$\widehat k^2=\sum_{i=1}^D\,\widehat k_i^2$. Then in D-dimensions the 
perturbative result (\ref{eq:wltrv}) reads
\begin{eqnarray} 
W_P(v)&=&\lambda_2 |v|^2 +\lambda_3 \Re v^3+
\frac{D-1}{L^2}\sum_{k_1k_2...k_D}\log\,
\Biggl [\biggl (\widehat k^2+\frac{g^2}{4}+\frac{3\gamma^2}{2}\biggr
)^3 \nonumber \\
&-&\frac{9\gamma^2}{2}
\biggl (\widehat k^2+\frac{g^2}{4}+\frac{3\gamma^2}{2}\biggr )
\biggl (\widehat k^2+\frac{g^2}{4}+3\,\gamma^2\biggr) |v|^2 \nonumber \\
&+&\frac{27\gamma^4}{2}\biggl (\widehat k^2+\frac{g^2}{4}+2\gamma^2\biggr
)\,\Re v^3
-\frac{81\gamma^6}{2^3}\, |v|^4 \Biggr ] .   \label{eq:Dwltrv}
\end{eqnarray}
Using it in dimension 3, we find that the phase structure 
represented in Fig.~\ref{fig:l2l3pd} persists, with similar values 
of $\lambda_2$ and $\lambda_3$ on the transition lines if the parameters
$\gamma^2$ and $g^2$ are kept the same. \\

{\bf Remark 2}

Heuristic arguments neglecting any entropic consideration 
help to understand the phase
diagram found, including the new phase III. The part $S_{U,V}$ of
the action is trivially minimized if the commutator $[U,V]$ vanishes
for any $U$, which implies $V\in Z_3$, i.e. in phase II. Conversely,
typically for $\lambda_2>0$ and $\lambda_3=0$, 
$S_V$ is minimized by $v$=0 (phase I) and there is 
a competition. A possible compromise is to require $V$ to
commute  with all the elements of an SU(2) subgroup of SU(3), that is to be
proportional to a $Z_2$ subgroup of $Z_3$: In diagonal form, 
$V=$diag$[\exp(i\theta),\,\exp(i\theta),\,\exp(-2i\theta)]$, up to colour
index permutations. In this subset $|v|$ is minimum for 
$\theta =(2\,n+1)\pi/3$, and  there $v=1/3\,\exp[i\,(2\,n+1)\pi/3]$ is
in phase III. \\

{\bf Remark 3}

The physical region for $v$ is limited by the condition that 
two (or three) eigenvalues of $V$ are equal. The points characterizing
phases II and III belong to this boundary, and the measure $\cD\,V$
vanishes there. This contributes a logarithmic repulsive potential on
the boundary. Its effect is  expected to
repell the values of $v$ in phases II and III slightly inside the physical 
region, and to change the exact position of the transitions lines.

\section{Summary}

We have investigated a model for Wilson lines
coupled to a $SU(3)$ gauge field in two dimensions.
This model was inspired by dimensional reduction of
three dimensional $SU(3)$ gauge theory. It is a 
$Z_3$ symmetric Higgs
model with the Wilson line acting as a Higgs field in the 
adjoint representation of the group. In perturbative dimensional
reduction, where the $Z_3$ symmetry is explicitly broken, 
the self couplings of the corresponding Higgs
field are determined by the perturbative integration
of the non static modes in the original theory. Here
we have left the selfcouplings free parameters, to be eventually
determined from the full model. 
The price paid is a loss in predictive power, but the model allowed us
to investigate in detail a non trivial phase diagram, which
includes the confinement-deconfinement transition of $QCD_3$.

In fact, the model exhibits three different phases. The order parameter is the
trace of the Wilson line, i.e. the Polyakov loop.
At low values of $\beta_2$ we find a confined phase,
where the Polyakov loop is distributed around zero. Above a
transition we find for $\lambda_2=0$ a deconfined phase, where
the Polyakov loop is near to an element of $Z_3$, one of the
third roots of unity. Keeping $\beta_2$ fixed in the deconfined
phase and letting $\lambda_2$ grow we find a first order phase
transition into a new phase, where the Polyakov loops are situated
near the boundary of phase space but with a phase half way in
between the third roots of unity. Here also, $Z_3$ symmetry
is spontaneously broken, but the Wilson line is  near to an element of
a $Z_2$ subgroup of SU(3).
If we increase $\lambda_2$ even more
we come into a confined phase again.

We also have measured the screening masses from the Polyakov
loop correlations. They become small near the deconfinement
phase transition. If one uses the screening mass to compare with
the full model, in order to find the values of the coupling constants
which correspond to dimensional reduction, one finds it to be in the
normal deconfined phase.

Considering this Wilson line model as a possible dimensionally reduced model of
three dimensional SU(3) gauge theory, and comparing it with usual perturbative 
dimensional reduction one finds that

- the model has explicit $Z_3$ symmetry in the action,
in contrast to the perturbatively reduced model, where the symmetry
is explicitely broken.

- it has a confinement-deconfinement transition in contrast to
the perturbatively reduced model

- at this deconfinement transition, which is in the same universality
class as full $QCD_3$, the screening masses should  go to zero.

- the reduction point in parameter space seems to lie in the physical phase,
in contrast
to the perturbative reduction, where it is in the metastable region
inside the Higgs phase

All these properties makes the Wilson line model an interesting
extension of dimensional reduction, if one wants to approach the
phase transition, while perturbative reduction can only be used
at temperatures higher than about two times the critical temperature.

There are several interesting questions that can be further
investigated.
In particular one would like to extend the numerical calculation to
three
dimensions, where one would have a dimensionally reduced model of
$(3+1)D$ QCD, i.e. the physical theory. Furthermore one 
should address the question if the new phase plays any role at
high temperature or high temperature and finite density QCD.
It is of course also important
to restrain the parameters, so as to get quantitative predictions
without parameters, as it is the case in perturbative dimensional reduction.

\section{Acknowledgments}

All the simulations were done on the PC--farm donated to P.B. by the
Alexander von Humbold Foundation. In 2003--2004, P.B. was partially supported 
by the KBN grant 2P03B 083 25. B.P. is grateful to the Service de physique 
th\'eorique
de Saclay (France) for support and kind hospitality. B.P. and P.B. also
thank DFG for support under the contract FOR 339/2-1.

\appendix

\section{Calculation of the Perturbative Polyakov Loops Potential \label{app:Wv}}

Here we derive the expressions \eqref{eigenm} of the eigenvalues of the
M-matrix, and the final expression of
the potential as a function of the Polyakov loop. 
The following two trace identities in SU(3) will be used extensively
\bea
\sum_{a}\,\tr (T^aXT^aY)=\frac{1}{2}\biggl (\tr X\,\tr Y-
\frac{1}{3}\,\tr (XY)\biggr ),  \label{id1}\\
\sum_{a}\,\tr (T^aX)\,\tr (T^aY)= \frac{1}{2}\biggl (
\tr (XY)-\frac{1}{3}\,\tr X\,\tr Y\biggr ). \label{id2} 
\eea
Recalling that $\tr T^a=0$, one first shows from its definition (\ref{mab}) 
that the matrix $M(V)$ is real and fulfills
\bea 
M^T(V)&=&M(V^\dagger) \nonumber\\
M(V^\dagger)\,M(V)&=&M(V)\,M(V^\dagger)=I, \label{idem1}\\
M^p(V)&=&M(V^p).\label{idem2}
\eea
Hence $M$ is orthogonal. In the matrix $Q$ of Eq. (\ref{qkm}), 
we factorize the matrix $K$ of Eq. (\ref{kij}), call $K_i$ its
eigenvalues and  write 
\bea 
\Tr \log Q&=&\Tr \log K+X,  \nonumber\\ 
\Tr \log K&=&\log\,K_1^8\,+\,\log\,K_2^8, \nonumber\\
X&=&-\sum_{p=1}^\infty\,\frac{\gamma^{2p}}{p\,2^p}\,
\Tr \biggl(K^{-p}\,\biggl [\,M(V)+M(V^\dagger)\,\biggr ]^p\biggr ). \label{qkx}
\eea
Note that $K^{-1}$ exists for any $\xi >0$. 
The Landau gauge is reached in the limit $\xi \to 0$ 
The trace involved in $X$ is the product of
$K_1^{-p}+K_2^{-p}$ by the trace $m_p$ in color space of $(M+M^\dagger)^p$,
\begin{equation}
m_p=\sum_{a=1}^8\,\biggl (\biggl [M(V)+M(V^\dagger)\biggr ]^p\biggr)_{aa}.
\end{equation}
In this expression, we expand $(M+M^\dagger)^p$ and apply 
(\ref{idem2}):
\begin{equation}
m_p=4\,\sum_{q=0}^p\,C_p^q\,\sum_{a,b=1}^8\,
\tr \biggl [T^a\,V^q\,T^b\,V^{\dagger q} \biggr ]\,\,
\tr \biggl [T^b\,V^{\dagger\,p-q}\,T^a\,V^{p-q} \biggr ].
\end{equation}
Repeated use of (\ref{id1}, \ref{id2}) to sum over color indices leads
to
\begin{equation}  \label {mvp}
m_p=\sum_{q=0}^p\,C_p^q\,\biggl (\tr  \biggl [V^q\,V^{\dagger p-q} \biggr ]\,\,
\tr \biggl[V^{\dagger q}\,V^{p-q} \biggr ]\,-1\biggr ).
\end{equation}
Let $V_\alpha \equiv \exp(i\,\theta_\alpha)$, $\alpha=1,2,3$ be the eigenvalues of $V$; those of $V^\dagger$ are $1/V_\alpha$  so that
\bea
m_p&=&\sum_{q=0}^p\, C_p^q\,\Biggl (\sum_{\alpha,\beta=1}^3\, 
\biggl (\frac {V_\alpha}{V_\beta}\biggr )^{2q-p}-1\Biggr )   \nonumber\\
&=&2\,\Biggl (2^p+\sum_{\alpha<\beta}\biggl (\frac{V_\alpha}{V_\beta}+
\frac{V_\beta}{V_\alpha}\biggr )^p \Biggl ). \label{mp} 
\eea

In Eqs. (\ref{qkx}), the resummation over p gives $X$, and $W$ follows.
Up to additive constants independent of $V$, $X$ and $W$ are then given by
\bea  
X&=&2\,\sum_{i=1}^2\,\log\Biggl [\,
\prod _{\alpha<\beta=1}^3
\,\Biggl (1-\frac{\gamma^2}{2\,K_i}\,
\biggl (\frac{V_\alpha}{V_\beta}+\frac{V_\beta}{V_\alpha}
\biggr )\Biggr )\Biggr ], \label{xw}\\
W(V)&=&\frac{1}{L^2}\,\sum_{k_1k_2}\,\sum_{i=1}^2\log\,\Biggl [\,
\prod _{\alpha<\beta=1}^3
\,\Biggl (K_i-\frac{\gamma^2}{2}
\biggl (\frac{V_\alpha}{V_\beta}+\frac{V_\beta}{V_\alpha}
\biggr )\Biggr )\Biggr ]. \label{wvvp} 
\eea
Eq. (\ref{xw}) proves that the eigenvalues of the symmetric part 
of $M$ are $\Re\, V_\alpha /V_\beta =cos(\theta_\alpha - \theta_\beta)$, 
as stated in \eqref{eigenm}. 
The next task is to express $W$ as a function of $\tr V$, and for this we 
have to evaluate the product
\begin{equation}
P=\prod_{\alpha <\beta=1}^3\,\biggl (1-\tau\,\frac{V_\alpha^2+V_\beta^2}
{V_\alpha\,V_\beta}\biggr )
\end{equation}
for  $\tau\equiv \gamma^2/(2\,K_i)$. Set $P=\sum_{n=0}^3 
(-1)^n\,c_n\,\tau^n$ and use $V_1\,V_2\,V_3=1$. We have 
$c_0=1$ and, $c.p.$ denoting the circular permutations of $\{1\,2\,3\}$,
\bea
c_1&=&(V_1^2+V_2^2)\,V_3\,+\, c.p. \nonumber\\
c_2&=&(V_1^2+V_2^2)\,V_3\,(V_2^2+V_3^2)\,V_1\,+\,c.p.\nonumber  \\
c_3&=&(V_1^2+V_2^2)\,(V_2^2+V_3^2)\,(V_3^2+V_1^2). \nonumber
\eea
Replacing systematically $V_\alpha^2+V_\beta^2$ by $\tr V^2-V_\gamma^2$
whenever $\alpha,\beta,\gamma$ are different, we find 
\bea 
c_1&=&\tr V^2\,\tr V\,-\,\tr V^3, \nonumber\\
c_2&=&\tr V^2\,\tr V\,+\,\tr V^{\dagger\,3}, \label{coefp} \\
c_3&=&\tr V^2\,\tr V^{\dagger\,2}\,-\,1. \nonumber
\eea
Now $\tr V^n$ for $n>1$ is a polynomial of degree n in $\tr V$ (for
general formulae, see Appendix A in \cite {us2}). In particular
\bea
\tr V^2&=&(\tr V)^2\,-\,2\,\tr V^\dagger, \nonumber\\
\tr V^3&=&(\tr V)^3\,-\,3\,\vert \tr V\vert ^2\,+\,3.\nonumber
\eea
Substituting these expressions into Eqs.(\ref{coefp}) and using
$v=1/3\,\Tr\,V$ provide the
desired result: Up to an additive constant, the perturbative part of the
Polyakov loop potential is
\bea
&&W(v)=\frac{1}{L^2}\,\sum_{k_1k_2}\,\sum_{i=1}^2\log
\Biggl [
(\frac{\gamma^2}{2}+K_i)^3 \nonumber\\
&&-\frac{9\gamma^2}{2}(\frac{\gamma^2}{2}+K_i)
(2\gamma^2+K_i)|v|^2 
+\frac{27\gamma^4}{2}(\gamma^2+K_i)\Re v^3 
-\frac{81\gamma^6}{2^3}|v|^4 \Biggr ], \label{wtrv}  
\eea
where the eigenvalues $K_i$ are
\begin{equation} 
K_1=\widehat k^2+\gamma^2+\frac{g^2}{4} \quad ; \quad  
K_2=\frac {\widehat k^2}{\xi }\,+\,\gamma^2+\frac{g^2}{4} \label{eigenk}
\end{equation}

\end{document}